\documentclass[prd,nofootinbib,showpacs,preprint,aps]{revtex4-1}
\usepackage{subfigure}
\usepackage{amsmath}
\usepackage{amsfonts}
\usepackage{amssymb}
\usepackage{graphicx}
\usepackage{advdate}

\usepackage[applemac]{inputenc}
\usepackage[T1]{fontenc}
\usepackage{amssymb}
\usepackage{amsfonts}
\usepackage{amsfonts} 
\usepackage{graphicx} 

\usepackage{bbm}
\def\beq{\begin{equation}}
\def\eeq{\end{equation}}
\def\bsp{\begin{split}}
\def\esp{\end{split}}
\def\bea{\begin{eqnarray}}
\def\eea{\end{eqnarray}}
\def\ba{\begin{array}}
\def\ea{\end{array}}
\def\nn{\nonumber \\}

\def\l.{\left.}
\def\r.{\right.}

\def\part{\partial}

\def\ie{{\it i.e. }}

\def\nn{\\ \nonumber}

\begin{document}
\preprint{UdeM-GPP-TH-26-308}
\preprint{arXiv:xxx.xxxxx}
\title{Reflecting Gravitons: The Graviton Laser and the Gertsenshtein effect.}
\author{Thomas Forget $^{1,3}$} 
\email{thomas.forget.1@umontreal.ca}
\author{M. B. Paranjape$^{1,2,3}$} 
\email{paranj@lps.umontreal.ca}
\author{Urjit Yajnik$^3$}
\email{urjit.yajnik@iitgn.ac.in}
\affiliation{$^1$Groupe de physique des particules, D\'epartement de physique,
Universit\'e de Montr\'eal,
C.P. 6128, succ. centre-ville, Montr\'eal, 
Qu\'ebec, Canada, H3C 3J7 }
\affiliation{$^2$Centre de recherche mathématiques and Institut Courtois}
\affiliation{$^3$Indian Institute of Technology Gandhinagar
Palaj, Gandhinagar - 382055,
Gujarat, India}
\begin{abstract}\baselineskip=18pt
Graviton lasers have been considered in the past, \cite{gl}, but practical terrestrial implementations appear infeasible. The absence of any known mechanism to reflect gravitons means that  it remains unclear how a graviton beam could be directed repeatedly through a putative lasing medium.
Astrophysical graviton lasing is still a possibilty as circular graviton orbits around blackholes afford the possibility of an arbitrarily long path length through the lasing medium of ultra-light dark matter \cite{bhgl,nhaxs}. In this essay, we consider the possibility of a graviton laser that could be constructed in a laboratory setting.  The graviton lasing medium could be one of many possible gravitating systems, of which we give three possible examples.  We calculate the possibility of reflecting the gravitons by using the conversion of gravitons into photons in an external magnetic field, the Gertsenshtein effect, \cite{Gertsenshtein1962}.  We may convert the gravitons to photons, then reflect the photons, then reconvert the photons into gravitons via the same effect, and then pass them through the graviton lasing medium.  With an identical apparatus on the other side, we can essentially extend the path length of the gravitons through the lasing medium as arbitrarily long as desired.

\vskip 1cm
\leftline{Essay written for the Gravity Research Foundation 2026 Awards for Essays on Gravitation.}
\leftline{March 31st, 2026}
\leftline{Corresponding author: M. B. Paranjape, paranj@lps.umontreal.ca}
\end{abstract}
\maketitle
\section{Introduction}     
The idea of a graviton laser, \cite{gl}, was first imagined in the context of ultra-cold neutrons bouncing on a table \cite{golub1991ultra} as described in the Q-bounce experiment.  The neutrons were organized in the vertical direction according to the Schr\"odinger equation in the earth's gravitational field, appropriately linearized, $V=m_N g z$, where $m_N$ is the neutron mass, $g$ is the gravitational acceleration and $z$ is the vertical distance above the table.  These neutrons could be used as a graviton lasing medium.\footnote{It is not essential that we use ultra-cold neutrons as the lasing medium.  Any quantum mechanical gravitating system will do.  For example, the papers, \cite{bhgl,nhaxs}, used ultra-light dark matter particles in quantum mechanical gravitationally bound states around black holes.  This is perhaps difficult to source in a laboratory supply store.  Alternatively, the mirrors at LIGO \cite{ligomirrors} are actually macroscopic objects that are in a superposition of the lowest 200 or so quantum mechanical states of a quantum harmonic oscillator. These could be used equally well as a lasing medium.}  A neutron in an excited state can decay  by the emission of a graviton.  If the graviton passes through a medium of excited neutron states (\ie an ensemble of neutrons that have a population inversion, more higher energy states are occupied than lower energy states) then the gravitons can produce stimulated emission, exactly as light does in an ordinary laser.  The problem is that the graviton laser has to be single pass, there are no natural ways of reflecting the gravitons so that they can pass through the lasing medium multiple times.  It is the content of this essay to point out that it is indeed possible to reflect gravitons through the following process.   Using the Gertsenshtein effect, \cite{Gertsenshtein1962}, first convert them to photons, then reflect the photons, re-convert them to gravitons, pass them through the graviton lasing medium and amplify them and have an identical apparatus on the other side to send them back.  
\begin{figure}[ht]
\centerline{\includegraphics[scale=.5]{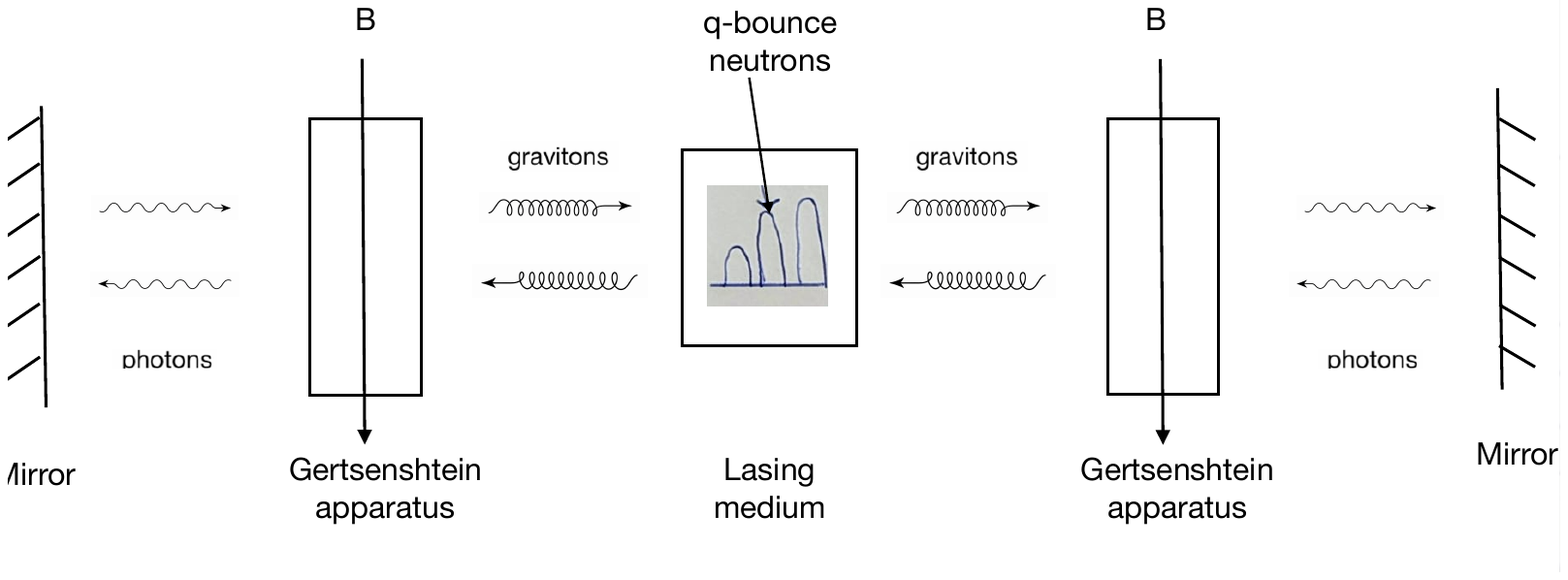}} \caption{\label{fig2} Graviton-photon laser with the Gertesenshtein effect. The lasing medium pictured is the ultra-cold neutrons bouncing on a table as in the Q-bounce experiment \cite{golub1991ultra}.}\label{fig2}
\end{figure}

\section{The Gertsenshtein effect}
In 1962, Gertsenshtein, \cite{Gertsenshtein1962},  predicted that gravitons passing through a background electromagnetic field could convert into photons.  The simple observation comes from the covariant Lagrangian for the electromagnetic field, interacting with gravity
\beq
{\cal L}= \sqrt{-g}\left(\frac{R}{8\pi\kappa }+ g^{\mu\nu}g^{\sigma\tau}F_{\mu\nu}F_{\sigma\tau}\right).
\eeq
\vfill\eject\noindent
Expanding the metric as $g^{\mu\nu}=\eta^{\mu\nu} +h^{\mu\nu}$  where $\eta^{\mu\nu} $ is the Minkowski metric we get
\[
\mathcal{L}_{\text{int}}
=
\kappa h^{\mu\nu}
\left(
F_\mu^{\,\,\,\alpha}F_{\nu\alpha}
-\frac{1}{4} \eta_{\mu\nu}F^{\alpha\beta}F_{\alpha\beta}
\right).
\]
Expanding further $F_{\mu\nu}=F^0_{\mu\nu}+f_{\mu\nu}$, where $F^0_{\mu\nu}$ is the background electromagnetic field that is taken to be constant, and $f_{\mu\nu}$ is the photon field,
gives the relevant, quadratic, cross term
\[
\mathcal{L}_{int.}=-{\cal H}_{int}
=2\kappa\left( h^{\mu\nu} F^{0\,\,\,\,\alpha}_{\,\,\,\,\mu} f_{\nu\alpha}
-\frac{1}{4} h^{\mu}{}_{\mu} F^{0\alpha\beta} f_{\alpha\beta}\right).
\]
Then classically, the graviton and photon mix as the quadratic part of the Lagrangian contains off-diaogonal terms.  Our aim is to calculate what happens if the external magnetic field is of finite extent.  In that case incoming gravitons will have a probability to turn into photons when they exit the magnetic field region.

The relevant part of the interaction Hamiltonian is actually time independent, showing that the interaction, if exisiting for all the time, would in fact change the definition of what the propagating fields/particles should be.  Using the expansion in terms of annihilation and creation operators for the gravitons
\beq
h^{\mu\nu}=\frac{1}{\sqrt V}\sum_{\vec q}\left(e^{-iq\cdot x}\epsilon^{\mu\nu} B(\vec q,\epsilon)+ e^{iq\cdot x}\epsilon^{*\mu\nu}B^\dagger(\vec q,\epsilon)\right)
\eeq
and the corresponding expansion for the photons
\beq
A_{\mu}=\frac{1}{\sqrt V}\sum_{\vec p}\left(e^{-ip\cdot x}\lambda_{\mu}A(\vec p,\lambda)+e^{ip\cdot x}\lambda^*_{\mu}A^\dagger(\vec p,\lambda)\right)
\eeq
where the momenta $q$ and $p$ are on shell, we find, writing  explicitly only the term that annihilates one graviton and creates one photon,
\beq
\int d^3 x {\cal H}_{int}=-2\kappa\sum_{\vec k, \lambda,\epsilon}B(\vec k,\epsilon)A^\dagger(\vec k,\lambda)\left(F^{0\,\,\,\,\alpha}_{\,\,\,\,\mu} \epsilon^{\mu\nu}(ik_\nu\lambda^*_{\alpha}-ik_\alpha\lambda^*_{\nu})-\frac{1}{4} F^{0\alpha\beta}\epsilon^{\mu}{}_{\mu}(ik_\alpha\lambda^*_{\beta}-ik_\beta\lambda^*_\alpha)\right)  +\cdots
\eeq
We note that the terms containing $B^\dagger A$ will also be time independent.  Therefore, the relevant part of the interaction Hamiltonian will mix the graviton states into the photon states in a time independent manner.

\section{Amplitude for mixing of gravitons and photons}
The amplitude for the absorption (or for the stimulated emission) of a graviton in a plane wave mode $\vec q,\,\,\epsilon^{\mu\nu}$,  and the corresponding emission (or absorption) of a photon with momentum vector $\vec p$ and polarization $\lambda$ is proportional to the matrix element  
\beq
{\cal M}=\int d^3 x \langle (M_{\vec p}+1, p^\nu, \lambda);(N_{\vec q}-1, q^\mu,\epsilon)|{\cal H}_{int}|(M_{\vec p}, p^\nu, \lambda);(N_{\vec q}, q^\mu,\epsilon)\rangle
\eeq
where a state $|(M_{\vec p}, p^\nu, \lambda);(N_{\vec q}, q^\mu,\epsilon)\rangle$ corresponds to $M_{\vec p}$ photons in the plane wave state $p^\nu, \lambda$ and $N_{\vec q}$ gravitons in a plane wave state $ q^\mu,\epsilon$ where $q^\mu$ and $p^\nu$ are on shell, \ie $c|\vec q |=q^0$, $c|\vec p |=p^0$ (there could be  many other occupied states in the system, but they are just spectators).   
The matrix element factorizes into the graviton matrix element and the photon matrix element, we have for generic annihilation and creation operators, 
\beq
\langle n+1, \vec p |a(\vec k)^\dagger|n,\vec p\rangle=\sqrt{n+1}\delta_{\vec k,\vec p}\,\, ,\quad \langle n-1, \vec p|a(\vec k)|n, \vec p\rangle=\sqrt{n}\delta_{\vec k,\vec p}
\eeq
hence
\bea
&~&{\cal M}=\nn
&~&=-2\kappa\sum_{\vec k, \bar\lambda,\bar\epsilon}\left[\left(F^{0\,\,\,\,\alpha}_{\,\,\,\,\mu} \bar\epsilon^{\mu\nu}(ik_\nu\bar\lambda^*_{\alpha}-ik_\alpha\bar\lambda^*_{\nu})-\frac{1}{4} F^{0\alpha\beta}\bar\epsilon^{\mu}{}_{\mu}(ik_\alpha\bar\lambda^*_{\beta}-ik_\beta\bar\lambda^*_\alpha)\right) \times\right.\nn
&~&\left.\times\langle (M_{\vec p}+1, p^\nu, \lambda);(N_{\vec q}-1, q^\mu,\epsilon)|B(\vec k,\bar\epsilon)A^\dagger(\vec k,\bar\lambda)|(M_{\vec p}, p^\nu, \lambda);(N_{\vec q}, q^\mu,\epsilon)\rangle\right]=\nn
&~&=-2\kappa\sum_{\vec k, \bar\lambda,\bar\epsilon}\left[\left(F^{0\,\,\,\,\alpha}_{\,\,\,\,\mu} \bar\epsilon^{\mu\nu}(ik_\nu\bar\lambda^*_{\alpha}-ik_\alpha\bar\lambda^*_{\nu})-\frac{1}{4} F^{0\alpha\beta}\bar\epsilon^{\mu}{}_{\mu}(ik_\alpha\bar\lambda^*_{\beta}-ik_\beta\bar\lambda^*_\alpha)\right) \sqrt{M_{\vec p}+1}\sqrt{N_{\vec q}}\,\, \delta_{\vec k,\vec p}\,\, \delta_{\vec k,\vec q}\,\,\delta_{\lambda\bar\lambda}\delta_{\epsilon\bar\epsilon}\right]\nn
&~&=-2\kappa\left(F^{0\,\,\,\,\alpha}_{\,\,\,\,\mu} \epsilon^{\mu\nu}(iq_\nu\lambda^*_{\alpha}-iq_\alpha\lambda^*_{\nu})-\frac{1}{4} F^{0\alpha\beta}\epsilon^{\mu}{}_{\mu}(iq_\alpha\lambda^*_{\beta}-iq_\beta\lambda^*_\alpha)\right) \sqrt{M_{\vec q}+1}\sqrt{N_{\vec q}}\,\, \delta_{\vec q,\vec p}\nn
&~&=-2\kappa\left(F^{0\,\,\,\,\alpha}_{\,\,\,\,\mu} \epsilon^{\mu\nu}(iq_\nu\lambda^*_{\alpha}-iq_\alpha\lambda^*_{\nu})\right) \sqrt{M_{\vec q}+1}\sqrt{N_{\vec q}}\,\, \delta_{\vec q,\vec p}
\eea
as the graviton is transverse, traceless.  
What this means is that the state $|(M_{\vec q}, q^\nu, \lambda);(N_{\vec q}, q^\mu,\epsilon)\rangle$ will mix with the state $|(M_{\vec q}+1, q^\nu, \lambda);(N_{\vec q}-1, q^\mu,\epsilon)\rangle$, and vice versa, so that while propagating in the presence of the external field $F_{\mu\nu}^0$, it is a linear combination of these states that will be the propagating degrees of freedom.   But since the photon and graviton states are degenerate in energy,  with energy eigenvalue given by $E_{M_{\vec q},N_{\vec q}}=\hbar c |\vec q|(M_{\vec q}+N_{\vec q})$, the overall effect will be simply that the two propagating states will be  two orthogonal linear combinations, that will be slightly  split in energy.  

Consider an external magnetic field, uniform and in the $z$ direction, so $F^0_{ij}=\epsilon_{ij3}B^0$, where $z$ is identified a the 3 direction and $B^0$ is the external magnetic field.  Furthermore, consider a graviton travelling in the $x$ direction and polarized in the $+$ mode, Then we find
\beq
{\cal M}=-2\kappa\left(\epsilon_{ij3}B^0 \epsilon^{i\nu}(iq_\nu\lambda^{*j}-iq^j\lambda^*_{\nu})\right) \sqrt{M_{\vec q}+1}\sqrt{N_{\vec q}}\,\, \delta_{\vec q,\vec p}
\eeq
where the three index $\epsilon_{ijk}$ is the Levi-Civita symbol.  Now using that $q^\mu=(c|q|,q,0,0)$ and $\epsilon^{22}=-\epsilon^{33}=1$ or the only non-vanishing components of the polarization, the first term vanishes and the second term gives, for a plane polarized photon $\lambda^*_2=1$
\beq
{\cal M}=2\kappa\left(\epsilon_{213}B^0 \epsilon^{22}(iq\lambda^*_{2})\right) \sqrt{M_{\vec q}+1}\sqrt{N_{\vec q^1}}\,\, \delta_{\vec q,\vec p}=-i2\kappa B^0 q \sqrt{M_{\vec q}+1}\sqrt{N_{\vec q}}\,\, \delta_{\vec q,\vec p}.\label{9}
\eeq
Thus an incoming state of just $N_{\vec q}$ gravitons and $M_{\vec q}$ photons, impinging on a region with a constant external magnetic field, the propagating degrees of freedom will be simple, orthogonal, linear combinations of the graviton/photon states, with slightly different energies.  As the graviton and photon are degenerate in energy before the magnetic field,  the interaction will simply form energy eigenstates by a $\pi/2$ rotation of the graviton and photon states into one another.  Indeed, from Eqn.\eqref{9}, 
\beq
\left(\left(\begin{array}{cc}E & 0 \\0 & E\end{array}\right)+(\Theta(x)-\Theta(x-L))\left(\begin{array}{cc}0& -i\alpha \\ i\alpha & 0\end{array}\right)\right)\left(\begin{array}{c}|\psi\rangle \\|\phi\rangle\end{array}\right)=\tilde E \left(\begin{array}{c}|\psi\rangle \\|\phi\rangle\end{array}\right) 
\eeq
where $\alpha=2\kappa B^0 q \sqrt{M_{\vec q}+1}\sqrt{N_{\vec q}}$, and  $\Theta(x)$ is the Heavyside function.  Thus for $x<0$ there is no change, nor for $x>L$.  But for $x\in (0,L)$ the eigenfunctions are a simple mix $|\psi_\pm\rangle=(|\psi\rangle \pm i |\phi\rangle)/\sqrt 2$, with energies $E_\pm=E\pm\alpha$.  Then if initially, the state is just $|\psi\rangle$, impinging into the magnetic field, it becomes the  linear combination $|\psi\rangle=(|\psi_+\rangle +|\psi_-\rangle)/\sqrt 2$.  This state evolves to $(e^{-i(E+\alpha)L/c}(|\psi_+\rangle +e^{-i(E-\alpha)L/c}|\psi_-\rangle)/\sqrt2$ as it moves at the speed of light through the magnetic field of width $L$.  Then when it exits, it presents itself as the combination  $e^{-iEL/c}\left(\left(\cos(\alpha L/c\right)|\psi\rangle - i\left(\sin(\alpha L/c)\right)|\phi\rangle\right)/\sqrt2$.  Therefore, an initial state $|\psi\rangle$ will mix into a linear combination of $|\psi\rangle$  and $|\phi\rangle$.  If the former represents the graviton and the latter the photon, we get a conversion to photons by
\beq
|\psi\rangle\to -ie^{-iEL/c} \left(\sin(\alpha L/c)\right)|\phi\rangle/\sqrt2 .
\eeq
Thus we get arbitrary conversion to photons depending on the value of $\alpha L/c$.  This factor is presumably small, as it is proportional to the gravitational coupling constant $\kappa$.   However it  contains $L$, the length of the regions containng the magnetic field;  $q$ the wave vector of the graviton/photon;  $\sqrt{M_{\vec q}+1}\sqrt{N_{\vec q}}$ which is essentialy the geometric mean of the number of gravitons and photons that are in the incoming state and $B^0$, the strength of the external magnetic field.  We can imagine circumstances that $\alpha L/c$ is of order one, even $\pi/2$ for maximal conversion.  

The Gertsenshtein apparatus can be rather complicated, with systems of mirrors to collect and remove any photons created, from the magnetic field, lest they convert back into gravitons, but we leave these details to the experimentalists.

\section{Graviton Lasing}
The centre point of the apparatus is the graviton lasing medium.  This could be any of a number of quantum mechanical systems which would interact with gravitons and participate in stimulated emmision of gravitons.  The amplitude/cross section for such a process seems to be universal, with the cross section independent of the masses of the participants in the quantum mechanical system, \cite{gl,bhgl}.   For example, for the case of the quantum bouncing neutrons, \cite{gl} we found
\beq
\sigma=\left(\frac{\hbar G}{c^3}\right)\frac{64\pi^2}{(\alpha_{n'}-\alpha_n)^6}
\eeq
where $\alpha_{n}$ and $\alpha_{n'}$ are the zeros of the Airy functions which characterize the wave functions of the neutrons in the earths gravitational field, which are pure numbers.  The dimensional factor in front, $\hbar G/c^3$,  is just the Planck area, and the numerical factor characterizing the states between which the transition occurs is completely independent of the masses involved, the mass of the neutron and the mass of the earth.  An analogous calculation concerning transitions between the level with quantum numbers $n=3,l=2,m=\pm 2$ and the ground state  of the Newtonian hydrogen atom, \cite{bhgl,ammp}, of ultra-light dark matter candidates in bound states around black holes, gives a somewhat more complicated, but similar result, 
\beq
\sigma =\frac{\hbar G}{c^3}\beta
\eeq
with 
\beq
\beta=\frac{81\pi^2}{32}\left(\sqrt{\frac{2}{15}}\frac{2^7\cdot5832}{3^{3}\cdot\sqrt{5!}}\sum_{n''=2}^\infty -n''^5\frac{(n''-3)^{n''-3}}{(n''+3)^{n''+3}}\frac{(n''-1)^{n''-1}}{(n''+1)^{n''+1}}\right)^2\approx 1.91.
\eeq
Again we find the Planck area multiplied by a pure number that characterizes the states between which the transitions occur.  We imagine this to be the general situation and it is surely due to the equivalence principle.

Numerically, the Planck area is incredibly small, $\hbar G/c^3\approx 10^{-70}$ metres$^2$.  The gain per unit length for passage through the lasing medium is given by the cross section multiplied by the number density of the lasing sites.  Hence to obtain significant amplification requires a very high density of the lasing sites, and/or a large number of passages through the lasing medium.  In actual photon lasers, both methods are utilised, when the gain is sufficiently high, a single pass through the lasing medium can suffice, while normally mirrors send the optical pulse back and forth through the lasing medium multiple times to obtain appropriate amplification.  It is this method that we seek to effect for the gravitons in conjunction with the Gertsenshtein effect.  
\section{Discussion}
Once the gravitons impinge on the Gertsenshtein apparatus, we need the gravitons to be efficiently converted to photons.  Any gravitons not converted to photons will not be reflected at the mirror, and will be effectively lost for lasing purposes.  

To obtain $\alpha L/c\approx \pi/2$ we need to examine the factors that make up $\alpha=2\kappa B^0 q \sqrt{M_{\vec q}+1}\sqrt{N_{\vec q}}$, in natural units ($\hbar=c=1$).  The gravitational coupling constant $\kappa$ is the very small parameter, $\kappa=8\pi/M_{\rm p}^2$ where $M_{\rm p}=\sqrt{\hbar c/G}\approx 1.22\times 10^{19}GeV$ is the Planck mass.  Now terrestrial magnetic fields can achieve field intensity of $\bar B\approx 100\,\, {\rm Tesla}$ under very special circumstances \cite{htmf1,htmf2}, which in natural units is $\bar B\approx 2\times 10^{-14}GeV^2$,  however we will adopt this value to get an optimistic rule of thumb.  Let it be said, in   astrophysical situations, say the fields around magnetars \cite{magnetars,magnetars2}, the magnetic field intensity can be as high as $10^{11}\,\, {\rm Tesla}$,  in natural units is $2\times 10^{-3}GeV^2$, which of course can give significant amplification in that situation.  For the wave vector $q$ we will take for comparison $\bar q\approx 2\times 10^{-8}GeV$ corresponding to a wavelength of $500$ nanometres.   For the length we take as comparison,  $\bar L=1$ metre,  which in natural units is $\bar L=5.07\times 10^{15} GeV^{-1}$, this gives 
\bea
\alpha L/c&=&(16\times 5.07\,\,\pi /(1.22)^2)\times 10^{-38-14-8+15}(q/\bar q)(B^0/\bar B)(L/\bar L) \sqrt{M_{\vec q}+1}\sqrt{N_{\vec q}}\nn
&\approx& 171.2\times 10^{-45}(q/\bar q)(B^0/\bar B)(L/\bar L) \sqrt{M_{\vec q}+1}\sqrt{N_{\vec q}}.
\eea
We note that the magnetic field factor ratio can change by many orders of magnitude.  Nevertheless, $\alpha L/c$ is rather small except for the factor which is the geometric mean of the number of photons and gravitons in the incoming state, which can be rather large, compensating for all the small coefficients, giving rise to negligible suppression  of the amplitude for graviton/photon conversion. 

For example, LIGO \cite{ligo} is able to detect gravitational waves, which must be actually states of gravitons.  The number of gravitons produced in a black hole merger is huge, $N_{\vec q}\approx 10^{78}$, with a terrestrial flux in the range of $10^{25}$ gravitons per square metre.  Therefore, conversion of large fluxes of gravitons to photons and back can be unsuppressed.  Therefore, if potentially large intial sources of gravitons can be produced, the Gertsenshtein effect can be significant, enough so that we can imagine almost all of the gravitons are converted to photons, then reflected from the mirror and then converted back to gravitons, to be amplified in the lasing medium through multiple round trips through the lasing medium.

It seems that the amplification in the lasing medium is really rather small, \cite{gl,bhgl}, and we do not expect that the gain is appreciable in a single pass.  In previous analyses, there was no understanding on how to implement graviton mirrors, normally, one cannot reflect gravitons.  However, using the Gertsenshtein effect, one can envisage that the gravitons to be amplified, can be converted into photons, reflected, reconverted into gravitons and then reamplified in each cycle.   If sufficiently many round trips can be produced we can imagine significant amplification.   It does require a detailed analysis of the potential loss of gravitons and photons in the apparatus to fully judge if the apparatus can be successful in actually amplifying the graviton flux.  In conclusion, the prospects for observing graviton lasing or constructing a corresponding apparatus remain uncertain, but they cannot be entirely ruled out given our current understanding, and further investigation into these possibilities remains a worthwhile endeavour.

\section{Acknowledgments}  We thank NSERC Canada for financial support. We thank, for their hospitality, the Physics Department of the Indian Insitute of Technology Gandhinagar and where this work was conceived and written up.  We thank Richard MacKenzie for a useful discussion.

\bibliographystyle{apsrev}
\bibliography{graviton-scatt-GRFEC2025-new-version}

\end{document}